 \documentclass [12pt,a4paper      ]{article}
\usepackage{graphics}
\usepackage{times}

\DeclareFontFamily{OT1}{times}{}
\DeclareFontShape {OT1}{times}{m }{n }{ <-> ptmr }{}
\DeclareFontShape {OT1}{times}{bx}{n }{ <-> ptmb }{}
\DeclareFontShape {OT1}{times}{m }{it}{ <-> ptmri}{}
\DeclareFontShape {OT1}{times}{bx}{it}{ <-> ptmbi}{}
\usepackage{amsmath}
\usepackage{amsfonts}
\usepackage{amssymb}
\usepackage{latexsym}
\newcommand{\CON}{\overline}  

\begin{document}

\title{\bf\vspace{-2.5cm} The physics of antimatter induced fusion
                          and thermonuclear explosions\footnote{Published {\bf in} G. Velarde and E. Minguez, eds.,  Proceedings of the 4th International  Conference on Emerging Nuclear Energy Systems, Madrid, June 30/July 4, 1986 (World Scientific, Singapore, 1987) 166--169.  For the the circumstances of the delivery of this paper, which was the first presentation at a scientific conference of the correct physical processes leading to the ignition of a large scale thermonuclear explosion using less than a few micrograms of antimatter as trigger, see Ref.~\cite{14}. See also Ref.~\cite{15}.  Since this paper and its extended version, Ref.~\cite{10}, have been published, many reports have confirmed their correctness, see, e.g., Refs.~\cite{16,17}.}}

\author{{\bf Andre Gsponer and Jean-Pierre Hurni}\\
\emph{Independent Scientific Research Institute}\\ 
\emph{Box 30, CH-1211 Geneva-12, Switzerland}\\
e-mail: isri@vtx.ch\\}

\date{ISRI-86-07.4 ~~ \today}

\maketitle

\begin{abstract}

The  possibility of using antihydrogen for igniting inertial confinement  fusion pellets or triggering large-scale thermonuclear explosions is investigated.  The number  of antiproton annihilations required to start a thermonuclear burn  wave in either $DT$ or $Li_2DT$ is found to be about $10^{21}/\kappa^2$,  where $\kappa$ is the compression factor  of  the fuel to be ignited.  We conclude that the financial  and  energy investments  needed to produce such amounts of antiprotons would confine applications of antimatter triggered thermonuclear devices to the military domain.

\end{abstract}

\section{Introduction}

Matter-antimatter  interaction  produces  more 
energy  per unit mass than any other means  of 
energy  production. For  example, proton-antiproton
  annihilation  releases  275  times 
more  energy in the form of kinetic energy  of 
charged  particles than nuclear fission or  $DT$ 
fusion.   This  energy  is released by  simple 
contact  of antimatter with matter so that, in 
principle,  no ignition energy is required  to 
start  the  reaction.   It  is  therefore  not 
surprising   that   the   concept   of   using 
antimatter  as  an energy source has  been  in 
scientific  literature for decades \cite{1,2}.

     Other  practical  applications of   antimatter
 are under consideration.  For  example, 
antimatter  propulsion  systems  \cite{3},    space 
based  power generators \cite{4},  directed  energy 
weapons \cite{4}, cancer therapy \cite{5}. Finally, both 
Edward  Teller \cite{6,7,8} and Andrei Sakharov  \cite{9}, 
the   key   scientists   in  charge   of   the 
development of the H-bomb in their  respective 
countries,  show in their published scientific 
works  a  big  interest  in  the  annihilation 
properties of antimatter,  the nuclear process 
that  after fission and fusion could lead to a 
third generation of nuclear bombs. 

     This   paper   is   a   summary   of    a 
comprehensive assessment of the feasibility of 
producing  large quantities of antiprotons and 
using  them for igniting  inertial-confinement 
fusion   pellets  or  triggering  large  scale 
thermonuclear explosions \cite{10}.

\section{Matter-antimatter annihilation}
 
When  a particle meets it's antiparticle  they 
annihilate  and the energy equivalent to their 
total  mass ($2mc^2$) is converted  into  various 
new particles and kinetic energy \cite{8}.   In the 
case  of proton-antiproton annihilation,  many 
different reaction channels are possible, each 
resulting  in the production of  a   different 
number  of charged and neutral  particles.   A 
 good  approximation is that three charged  and 
two neutral pions are produced on the average.  
Since   neutral   pions  quickly  decay   into 
photons,  the typical $p\CON{p}$ annihilation  process 
is as follows: 
\begin{equation}\label{EQ-1}
   p + \CON{p} \rightarrow 3 ~ \pi^\pm + 4 ~ \gamma ,
\end{equation}
where $E_\pi^\pm = 236$~MeV and $E_\gamma = 187$~MeV. 
An  antiproton  can  also  annihilate  with  a 
neutron,   in  which  case  mostly  pions  are 
produced again,  in numbers,  on the  average, 
similar to $p\CON{p}$ annihilation. 

     Antiprotons,  antineutrons  and positrons 
can  combine to  form  antinuclei,  antiatoms, 
antimolecules.   Annihilation  occurs when the 
two kinds of matter come sufficiently close to 
one other.   Even at some distance,  a neutral 
atom and  a neutral antiatom will attract each 
other  by  van  der Waals forces  \cite{8}.   As  a 
consequence,   storage   of \emph{antiatoms} in   a 
container  made  of matter  is  impossible  in 
general.   However, there may exist metastable 
states of \emph{antiprotons} in normal matter \cite{11}.

\section{Plasma heating with antiprotons}
 
When  a  $\CON{p}$ annihilates in a  hydrogen  plasma, 
essentially  all  the annihilation  energy  is 
radiated  in the form of very energetic  pions 
and photons.  At solid hydrogen densities, the 
mean free path of the 187~MeV photons is 25~m, 
so  that  they  will not loose energy  in  the 
plasma.   However,  the three 236~MeV  charged 
pions  will  loose energy by multiple  Coulomb 
interactions  with  the electrons  at  a  rate 
approximately given by: $dE/dx = 0.52$~MeV/cm in 
solid $H_2$ or $DT$ and 2.06~MeV/cm in $Li_2DT$.
 
     If  we now assume that annihilation takes 
place  at the center of a sphere,  the  energy $dW$ 
deposited  within a radius $R=1$~cm is  only   
1.5~MeV  out  of the total 1876~MeV  annihilation 
energy.   There  are however several ways   to 
improve  energy  deposition,  and thus  plasma 
 heating.   Firstly,  the fuel to be heated may 
be compressed by a factor $\kappa$,  $dE/dx$ will  then 
be multiplied by $\kappa$,  and thus $dW$ by $\kappa^{2/3}$.  But 
compression requires energy.  Secondly,  fuels 
such  as $Li_2DT$,  which contain more electrons, 
have a proportionally larger $dE/dx$.   However, 
their  thermonuclear ignition  temperature  is 
also higher.   Finally,  annihilation may take 
place with a nucleus. 
 
     When  a $\CON{p}$ annihilates with a nucleon from 
a nucleus,  because of the Fermi motion of the 
annihilated nucleon,  the nucleus will  recoil 
with an energy of about 20 MeV.   Furthermore, 
each   of  the  5  annihilation  pions  has  a 
probability of colliding with the rest of  the 
nucleus.   Hence,  the  average  total  energy 
deposition in a sphere is 
\begin{equation}\label{EQ-2}
 dW = \nu \frac{dE}{dx} R + \epsilon ,
\end{equation}
where  $\nu = 3$ is the number of charged pions and    
$\epsilon$ the local energy deposition by the recoiling 
nucleus    and   the   various    pion-nucleus 
interaction debris. 

     In  the case of $\CON{p}$ annihilation with  
deuterium  or tritium  $\epsilon$ is approximately 12~MeV 
on the average,  about half of the Fermi energy.   
With  heavy nuclei there have been  many 
theoretical  speculations  in the  absence  of 
measurements.   The  first of these was introduced  
by Duerr and Teller \cite{7}, who  speculated 
that  an antiproton would find a  very  strong 
(900~MeV)  attractive potential when  getting 
close to a nucleus.   More recently \cite{12},  Los 
Alamos  scientists have calculated that annihilation  
in carbon would result in the  local 
energy  deposition of about 100~MeV.   Recent 
measurements  at CERN show that it is in  fact 
only  33~MeV in carbon \cite{5}.   Low energy  $\CON{p}$'s  
annihilate  mostly at the surface  of  nuclei, 
and  thus  local energy deposition  follows  a 
$A^{2/3}$ dependence on atomic weight.   In effect, 
the   CERN   data  is  compatible   with   the 
expression : 
\begin{equation}\label{EQ-3}
    \epsilon \approx 6.4 A^{2/3}  ~~~ \text{[MeV]}.
\end{equation}
Hence,  for $\CON{p}$ annihilation in $H_2$, $DT$ or $Li_2DT$,       
$\nu$  is  always  about 3 and $\epsilon$ is  approximately 
equal to 0, 12 or 22 MeV respectively. 
 
\section{Thermonuclear burn of a particle-antiparticle plasma}
 
A matter-antimatter plasma is obtained if some 
initially stable particle-antiparticle mixture 
is suddenly ignited.  The annihilation rate of 
two interacting species, with number densities 
$n$ and $\CON{n}$, is 
\begin{equation}\label{EQ-4}
        \frac{d\CON{n}}{dt} = - n \CON{n} \langle \sigma v \rangle ,
\end{equation}
where  $\langle \sigma v \rangle$ is the annihilation reaction rate 
averaged over the Maxwell distribution. 
 
\begin{figure}
\begin{center}
\resizebox{12cm}{!}{ \includegraphics{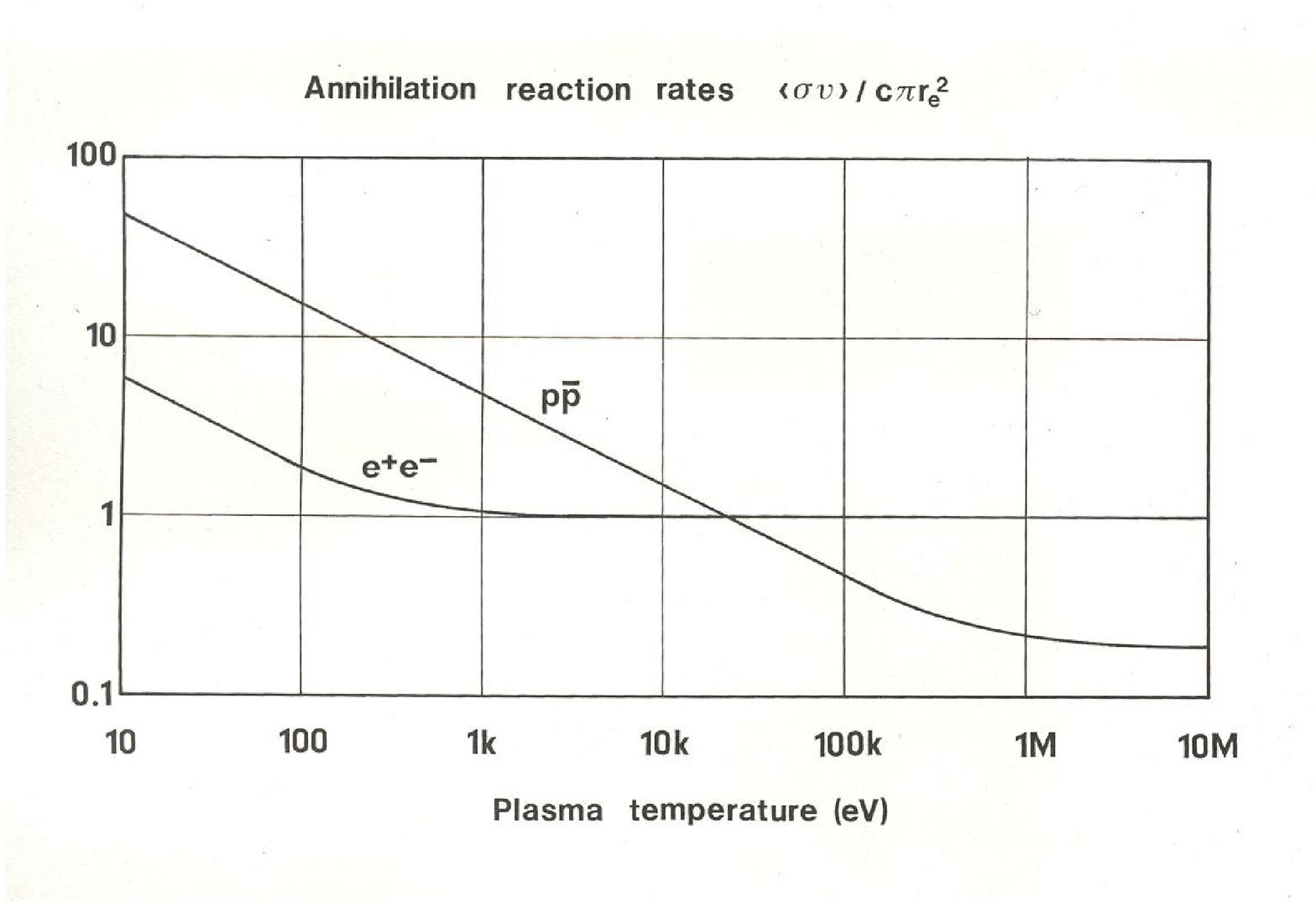}}
\end{center}
\caption[Annihilation reaction rates]{Electron-positron and proton-antiproton  annihilation reaction rates averaged over the Maxwell velocity distribution.}
\end{figure}

In a $H - \CON{H}$ plasma,  equation \eqref{EQ-4} holds for  both 
protons  and  electrons  with  $n = \CON{n}  =  n_0  = \rho \, \mathcal{N}_A/2$
initially. Hence, for a given temperature 
\begin{equation}\label{EQ-5}
     n = \frac{n_0}{1+t/\tau}  ~~~  ~~~ \text{with} ~~~ ~~~
  \tau =\frac{2}{n_0\langle \sigma v \rangle} .
\end{equation}
If  we assume $T = 20$~keV,  $\langle \sigma v \rangle$ is  approximately
  the  same for both $e^+ e^-$ and $p\CON{p}$  
annihilation.   Thus  the electron and the  proton 
populations deplete  at the same rate,  with a  
time constant  of  5~ns  for   $\rho = 0.07$~g/cm$^3$.

\section{Annihilation in a matter-antimatter boundary layer}
 
When  matter and antimatter come into contact, 
annihilation   primarily  takes  place  in   a 
boundary   layer   in  which   particles   and 
antiparticles are mixing.   The thickness   of 
this  matter-antimatter plasma is of the order 
of  the  antimatter mean-free-path in  matter, 
i.e.,  $(3n\sigma)^{-1}$.   A first approximation,  
assuming  that  whenever an  antiparticle  
penetrates  into  the boundary layer it  instantly 
annihilates,   is  an  annihilation  rate  per 
element  area  given by the  total  number  of 
antiparticles impinging on that surface.  From 
the Maxwell velocity distribution one gets 
\begin{equation}\label{EQ-6}
       \frac{d\CON{N}}{dS\,dt} = -\CON{n}c \sqrt{\frac{k T}{2\pi mc^2}}  .
\end{equation}
The $e^+$ annihilation rate is thus  $\sqrt{m_p/m_e}  \approx  43$ 
times the $\CON{p}$ annihilation rate.  However, since 
the $\CON{H}$ plasma Debye length is much smaller than 
the  boundary layer thickness,  plasma  charge 
neutrality  insures  that the antimatter  flow 
rate is determined by the slowest annihilation 
rate.  Therefore,  if  $\CON{H}$'s interact  with  the 
walls of a closed cavity, annihilation results 
in  an  overall  decrease  of  the  antimatter 
density within the cavity.
 
     Let  us now take the case of a sphere  of 
solid  antihydrogen  that is suddenly  put  in 
contact  with a collapsing spherical shell  of 
compressed $DT$.  To solve Eq.~\eqref{EQ-6} one has to 
calculate  the  increase in the $\CON{H}$  plasma  internal 
energy by the pions and other particles from $\CON{p}$ 
annihilation in the surrounding $DT$: 
\begin{equation}\label{EQ-7}
    dW = -dN \frac{1}{2}
          \Bigl( \nu \frac{dE}{dx} + \frac{\epsilon}{\lambda} \Bigr)
           \frac{4R}{\pi} \frac{N}{N_0} ,
\end{equation}
where $\lambda  = 3$~cm is the approximate range of the  
20~MeV recoil protons from $\CON{p}$ annihilation  in 
$DT$,  and $N$ (initially equal to $N_0$) the  number 
of $\CON{H}$ atoms.  For hydrogen  $dW = 3 N k dT$, we get 
a system of equations for the $\CON{H}$ plasma density 
and  temperature.   If  annihilation  is  much 
faster  than  the collapse of  the  cavity  ($R$ 
constant) the solution of Eqs.~\eqref{EQ-6} and \eqref{EQ-7} is 
\begin{equation}\label{EQ-8}
    T= T_1 \tanh^2 (t/\tau_a) ~~~ ~~~ \text{and} ~~~ ~~~
    N = N_0 \Bigl(1 -  \tanh^2 (t/\tau_a) \Bigr)    .
\end{equation}
For $N_0 = 10^{18}$,  which corresponds to $R =  0.02$~cm,  
we  find  $T_1 = 19$~keV and $\tau_a  = 0.25$~ns. 
Thus, in about  $2\tau_a  = 0.5$~ns, over 90\% of the 
antihydrogen  in  the sphere  is  annihilated.  
This  time  constant is  compatible  with  the 
requirements  of instantaneous  thermalization 
and inertial confinement of the plasma.

\begin{figure}
\begin{center}
\resizebox{8cm}{!}{ \includegraphics{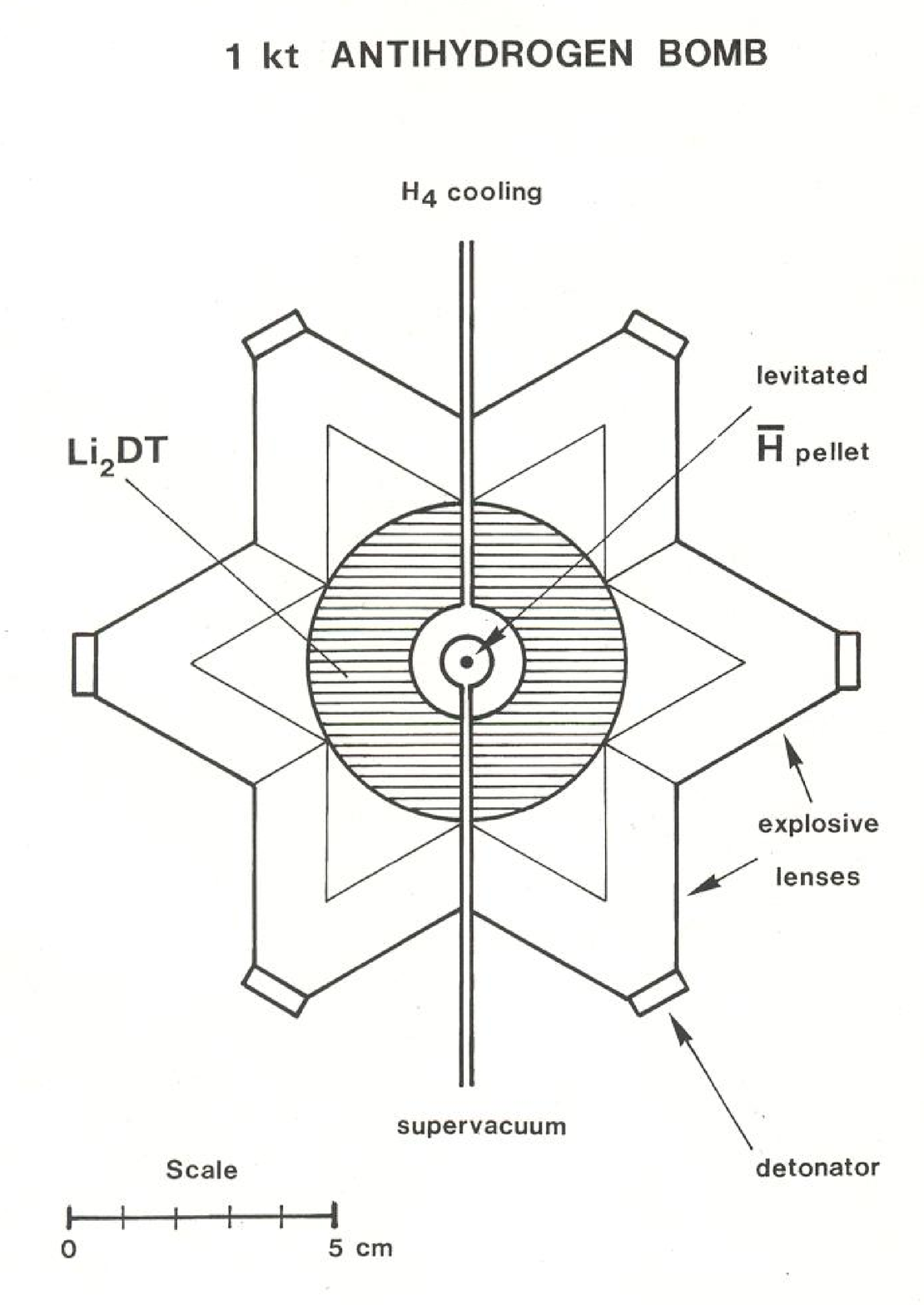}}
\end{center}
\caption[Antimatter bomb]{In  the  configuration for a 1~kt  
antimatter  bomb shown above,   one microgram of 
antihydrogen  in a microcryostat is  levitated 
at  the  center  of  a  100~g  $Li_2DT$  sphere.  
Implosion  of the $Li_2DT$ by means  of  chemical 
explosives  brings the thermonuclear fuel into 
contact  with the  antihydrogen.   The  energy 
release  by  annihilation  is fast  enough  to 
trigger  an outgoing thermonuclear  detonation 
wave which burns the $Li_2DT$.   Depending on the 
amount   of   compression  by   the   chemical 
explosives,  the  device  operates as a  1~kt 
neutron bomb (ERW --- Enhanced Radiation Warhead) 
or a 1~kt blast bomb (RRR -- Reduced Residual radioactivity).  
In either case,  the antimatter bomb will have 
very   reduced   radioactive   fallout    and 
electromagnetic pulse effects.
     From   the   point   of  view   of   
non-proliferation  of nuclear  weapons,  the  fact 
that     antimatter-triggered    thermonuclear 
weapons    will   have    extremely    reduced 
radioactive  fallout,  even for ground bursts, 
is  an important  consideration.   Since  such 
explosives  may  be  advocated  for  "peaceful 
nuclear   explosions,"    the   current   
non-proliferation regime is being threatened by the 
growing  spread  of  high  energy  accelerator 
technologies \cite{13}.  Moreover, from a strategic 
point   of  view,   the  possible  advent   of 
extremely   compact   and  essentially   clean  
nuclear  weapons  would  further  diffuse  the 
distinction between low-yield nuclear  weapons 
and conventional explosives.} 
\end{figure}

\section{Antiproton triggered thermonuclear detonation wave}
 
The   most   efficient   way  to   trigger   a 
thermonuclear explosion is probably to start a 
thermonuclear  detonation  wave  in  $Li_2DT$  by 
collapsing a hollow sphere of that material on 
a tiny spherical pellet of solid antihydrogen. 

     In   the  spark  model  of  thermonuclear 
ignition,  an  outgoing  spherical  detonation 
wave  starts  if :  (a) a critical  amount  of 
energy  $E_c$ is deposited in the center  of  the 
sphere  (the  "spark" region) and (b)  if  the 
temperature  within this volume is higher than 
a    critical   temperature   $T_c$.     Without 
compression,   one  has  $E_c = 5 \times 10^{25}$~keV   and  
$T_c  = 4$~keV  for  solid $DT$ ,   and $E_c = 3 \times 10^{26}$~keV 
and $T_c = 13.6$~keV for  $Li_2DT$.   However, 
for   a  compressed  thermonuclear   fuel   at 
temperature $T_c$,  the critical energy decreases 
with the square of the compression factor $\kappa$. 

     The number $N$ of $\CON{p}$ annihilations necessary 
to  induce  a thermonuclear burn wave  can  be 
estimated by supposing that annihilation takes 
place  at  the  center  of the  sphere  to  be 
ignited.   Thus,  from equation \eqref{EQ-2}, condition 
(a) is satisfied if 
\begin{equation}\label{EQ-9}
  E_c/\kappa^2 = N \Bigl( \nu \frac{dE}{dx} \kappa R_s + \epsilon \Bigr) .
\end{equation}
Since the pions originate from the center, the 
temperature   in   the  fuel  goes  as   $1/r^2$.  
Therefore,  for  simplicity,  we require  that 
condition  (b)  is satisfied for  the  average 
temperature within the critical volume.  Thus 
\begin{equation}\label{EQ-10}
E_c/\kappa^2 = \frac{3}{2} \frac{z}{a} \kappa \rho N \frac{4\pi}{3} R_s^3 kT_c ,
\end{equation}
where $z$ and $a$ are respectively equal to 2  and 
2.5 for $DT$,  and 6 and 9.5 for  $Li_2DT$.   Taking     
$\kappa =  30$,  a  modest  compression  factor,  and  
solving Eqs.~\eqref{EQ-9} and \eqref{EQ-10} for $N$ and the spark radius $R_s$, 
one finds $N = 3 \times 10^{18}$ and $R_s = 0.09$~cm for $DT$, 
and $N = 6 \times 10^{18}$ and $R_s = 0.07$~cm for  $Li_2DT$.  
However,  because  of some of the  simplifying 
assumptions   made,   these  results  may   be 
somewhat pessimistic.   Hence,  we will assume 
that $10^{18}$ $\CON{p}$'s are sufficient to trigger  the 
thermonuclear  explosion of compressed  $DT$ or 
$Li_2DT$ pellets. 

     For  thermonuclear explosions in  the  kiloton 
range,  chemical  explosives  may be  used  to 
implode  the  $Li_2DT$  shells.   For  low  yield 
explosions  such as in X-ray laser pumping  or 
ICF, compression factors higher than 30 can be 
achieved using magnetic compression,  beams or 
other techniques.  However, antiproton induced 
fusion  will remain an attractive  alternative 
to  normal ICF only if the compression  factor 
is kept relatively small,  i.e., less than 300, 
giving a number of $\CON{p}$'s of the order of $10^{16}$.

\section{Discussion}
 
  The   production  of  $10^{16}$  $\CON{p}$'s   for   each 
antimatter  triggered ICF pellet would require 
an energy investment of at least $10^4$~MJ  \cite{10}.  
It will therefore be very difficult to achieve 
energy break-even in power generating  reactors 
using annihilation techniques.   Moreover, the 
technologies   for  producing  $\CON{p}$'s  with  high 
energy accelerator systems,  and the means for 
manipulating and storing sizable amounts of  $\CON{H}$ 
are  extremely complicated.   For instance,  a  
plant  of  the size required  to  produce  the 
antimatter  needed for one thermonuclear  bomb 
trigger a day ($10^{-6}$g of $\CON{H}$ or $10^{18}$ $\CON{H}$ atoms per 
day)  could consist of several 10's of accelerators 
and storage rings,  and could require 
as many as several large nuclear power  plants 
to  supply the electricity \cite{10}.   A study  by 
the  RAND Corporation gives a cost estimate of 
\$500 to 1000  million  for  a  prototype  factory 
providing 10 to 100~micrograms, and \$5 to 15  billion 
for  a full production factory with an  output 
of about 10~mg per year \cite{4}. As a consequence, 
civilian applications of antimatter for  power 
production are very unlikely. 

     Directed  energy weapons applications may 
include the triggering of thermonuclear plasma 
jets,  and  X-ray or gamma-ray laser  pumping.  
In  the  event  of a  comprehensive  test  ban 
treaty,  antimatter would provide a means  for 
inducing  laboratory  and small scale  thermonuclear 
explosions  in a  yield  range  which 
cannot   easily  be  covered   by  underground 
explosions or classical ICF systems \cite{13}.   Of 
course,  many technical problems will have  to 
be solved \cite{10}.  In particular, the levitation of a 
frozen  $\CON{H}$ pellet  within a  1~mm  diameter 
cryostat   at   the   heart   of   a   complex 
thermonuclear device is a tremendous challenge 
for  materials microtechnology.   However,  if 
metastable  states  of  $\CON{p}$'s  in  $Li-$, $Be-$ or 
possibly $C-DT$ compounds are  discovered,  much 
simpler designs could be considered.


\begin{thebibliography}{999}


\bibitem{1} Yu.  D.  Prokoshkin: \emph{Particles of antimatter}.
 Die Naturwissenschaften {\bf 59} (1972) 282--284. 
N.A.   Vlasov:   \emph{Annihilation   as  an  energy 
process}. Sov. At. Energy {\bf 44} (1978) 40--45. 
 
\bibitem{2} H. Hora:  Laser interactions and related 
plasma phenomena {\bf 3B} (1974) 833.  
H. Hora:  Laser  plasmas  and  nuclear  energy, 
Plenum Press, New York (1975) 85--86, 97. 
 
\bibitem{3}  Special  issue  of the J.  of  the  Brit. 
Interplanetary Soc.  on antimatter propulsion. 
JBIS {\bf 35} (1982) 387-424. 
R.L.  Forward: \emph{Making and storing antihydrogen 
for  propulsion},  Workshop on the Design of  a 
Low-Energy  Antimatter  Facility in  the  USA, 
University of Wisconsin, October 3-5 (1985). 
 
\bibitem{4} B.W.  Augenstein:  \emph{Concepts, problems, and 
opportunities for use of annihilation  energy.  
Prepared for the United States Air Force}, RAND 
Note N--2302--AF/RC, June (1985) 61 pp. 
 
\bibitem{5} A.N.  Sullivan:  \emph{A measurement of the local 
energy  deposition  by antiprotons  coming  to 
rest  in tissue-like material}.  CERN Preprint 
TIS--RP/149/CF (22 April 1985). Phys. Med. Biol. {\bf 30} (1985) 1297--1303.
 
\bibitem{6} E.  Fermi and E.  Teller:  \emph{The capture  of 
negative  mesotron  in  matter}.   Phys.   Rev. 
{\bf 72} (1947) 399--408. 
 
\bibitem{7} H.P.  Duerr and E.  Teller: \emph{Interaction of 
antiprotons with nuclear  fields}.  Phys.  Rev. 
{\bf 101} (1956) 494--495. 
 
\bibitem{8}  D.L.   Morgan  and  V.W.  Hugues:  \emph{Atomic 
processes   involved   in    matter-antimatter 
annihilation}.  Phys.  Rev.  {\bf D2} (1970) 1389--1399. 
For E. Teller see page 1399.\\ 
D.L.   Morgan,   V.W.   Hughes:  \emph{Atom-antiatom 
interactions}. Phys. Rev. {\bf A7} (1973) 1811--1825.   
 
\bibitem{9}  A.  Sakharov:  Collected scientific work, 
Marcel Dekker Inc., New York (1982). 
 
\bibitem{10} A.  Gsponer and J.-P.  Hurni:  \emph{Antimatter 
induced  fusion and thermonuclear  explosions}.  Atomkernenergie $\cdot$ Kerntechnik  (Independent Journal on Energy Systems and Radiation) {\bf 49} (1987) 198--203.   e-print arXiv:physics/0507125 available in PDF format at\\ \underline{http://arXiv.org/pdf/physics/0507125}
 
\bibitem{11} M.V.  Hynes: \emph{Physics with low temperature 
antiprotons}, {\bf in} Physics with antiprotons  at 
LEAR  in  the ACOL era,  Editions  frontieres, 
Gif-sur-Yvette, France (1985) 657--664. 
 
\bibitem{12}   M.R.   Clover   et  al.:   \emph{Low   energy 
antiproton-nucleus  interactions}.  Phys.  Rev. 
{\bf C26} (1982) 2138--2151. 
 
\bibitem{13}  A.  Gsponer, B. Jasani, and S. Sahin:   \emph{Emerging  nuclear 
energy     systems    and    nuclear    weapon  
proliferation}.    Atomkernenergie $\cdot$ Kerntechnik  (Independent Journal on Energy Systems and Radiation) {\bf 43} (1983) 169--174.

\bibitem{14} A. Gsponer et J.-P. Hurni: \emph{Les armes \`a antimati\`ere}.  La Recherche {\bf 17} (novembre 1986) 1440--1443. English translations: \emph{Antimatter weapons}. The World Scientist (New Delhi, India, 1987) 74--77, and Bulletin of Peace Proposals {\bf 19} (1988) 444--450. e-print arXiv:physics/0507132 available in PDF format at\\ \underline{http://arXiv.org/pdf/physics/0507132}

\bibitem{15} A. Gsponer et J.-P. Hurni: \emph{Antimatter underestimated}.  Nature {\bf 325} (26 February 1987) 754. e-print arXiv:physics/0507139 available in PDF format at\\ \underline{http://arXiv.org/pdf/physics/0507139}

\bibitem{16} L.J. Perkins, C.D. Orth, and M. Tabak: \emph{On the Utility of Antiprotons as Drivers for Inertial Confinement Fusion}. Nucl. Fusion {\bf 44} (2004) 1097--1117. Available in PDF format at\\ \underline{http://cui.unige.ch/isi/sscr/phys/Perkins-Ort-Tabak.pdf}

\bibitem{17} M. Shmatov: \emph{The typical number of antiprotons necessary to heat the hot spot in D-T fuel doped with U}. JBIS {\bf 58} (2005) 74--81. Available in PDF format at\\ \underline{http://cui.unige.ch/isi/sscr/phys/Shmatov.pdf}


\end{thebibliography}
\end{document}